\documentclass[12pt]{article}
\usepackage{amssymb}

\def\IK{\relax{\rm l\kern-.18 em K}}
\def\IL{\relax{\rm I\kern-.18 em L}}

\def\ii{{\rm i}\,}

\def\pd#1#2{\frac{\partial #1}{\partial#2}}

\def\frac#1#2{{#1\over #2}}

\def\ptos{\leaders\hbox to 2mm{\hfil{.}\hfil}\hfill}
\def\\{\hfill\break}
\def\matriz#1#2{\left( \begin{array}{#1} #2 \end{array}\right) }
\def\braket#1#2{\langle#1\mathbin\vert#2\rangle} 
\def\Hil{{\cal H}}
\def\bra#1{\langle\,#1\!\mid} 
\def\ket#1{\mid#1\,\rangle}
\def\PH{\mathbb{P}\cal H}

\font\tenfrak=eufm10  \font\sevenfrak=eufm7  \font\fivefrak=eufm5
\newfam\frakfam
\textfont\frakfam=\tenfrak
\scriptfont\frakfam=\sevenfrak
\scriptscriptfont\frakfam=\fivefrak

\font\tengoth=eufm10 scaled\magstep1 \font\sevengoth=eufm7 \font\fivegoth=eufm5
\newfam\gothfam
\textfont\gothfam=\tengoth\scriptfont\gothfam=\sevengoth
  \scriptscriptfont\gothfam=\fivegoth

\def\today{\ifcase\month\or
   January\or February\or March\or April\or May\or June\or
   July\or August\or September\or October\or November\or December\fi
   \space\number\day, \number\year}

\title{Geometrical description of algebraic structures: Applications to Quantum
  Mechanics}

\author{J.F. Cari\~nena$^{a)}$, A. Ibort$^{b)}$, G. Marmo$^{c)}$ and G. Morandi$^{d)}$ }
\date{}
\begin{document}
\maketitle
\centerline{\footnotesize $^{a)}$ Departamento de F\'{\i}sica Te\'orica, Universidad de Zaragoza,}

\centerline{\footnotesize 50009 Zaragoza, Spain}
\centerline{\footnotesize jfc@unizar.es}
\medskip
\centerline{\footnotesize $^{b)}$
Departamento de Matem\'aticas,  Universidad Carlos   III de Madrid,}

\centerline{\footnotesize  Avda. de la Universidad 30, 28911 Legan\'es, Spain}

\centerline{\footnotesize albertoi@math.uc3m.es}

\medskip

\centerline{\footnotesize $^{c)}$
Dipartimento  di Scienze Fisiche,             
Universit\'a Federico II di Napoli e Sezione INFN di Napoli,}
\centerline{\footnotesize Via Cintia, 80125,
Napoli, Italy}

\centerline{\footnotesize marmo@na.infn.it}\medskip

\centerline{\footnotesize $^{d)}$
Dipartimento di Fisica, Universit\'a  di Bologna e Sezione INFN di Bologna}

\centerline{\footnotesize 6/2 viale B. Pichat, I-40127 Bologna, Italy}

\centerline{\footnotesize Giuseppe.Morandi@bo.infn.it}

\bigskip

\begin{abstract}  Geometrization of physical theories have always played an important role
in their analysis and development.  In this contribution we discuss various aspects concerning
the geometrization of physical theories: from classical mechanics to quantum mechanics.
We will concentrate our attention into quantum theories and we will show how to use in a systematic
way the transition from algebraic to geometrical structures to explore their geometry, mainly its Jordan-Lie structure. 
\end{abstract}

\bigskip

{\it Keywords:}{\enskip}
Symplectic, Poisson, connection, Hermitean

\medskip

PACS codes: 02.40.Yy; 03.65.Ca; 45.20.Jj 

\vfill\eject

\hfill ``{\it ... I realized that the foundations of geometry have physical relevance}.'' 

\hfill {\small A. Einstein letter to M. Besso, March 26, 1912.}

\medskip

\hfill ``{\it Tensor calculus knows physics better than most physicists}'' 

\hfill {\small L. Brillouin.}

\section{ Introduction}
The two quotations above should make clear why we would like to privilege a
geometrical description of physical systems.   The geometrical description of physical systems 
uses more general objects than
the traditional Euclidean spaces:  differentiable manifolds, sometimes
endowed with particular structures.
 After the formulation of General Relativity by means  of the (pseudo-) Riemannian
 geometry,
it is accepted without any doubt that the equations used to describe physical
systems should be written in tensorial form. For instance, we may indeed consider classical
Gauge Theories as the Theory of Connections.

Quantum theories, due to the superposition
rule, are always formulated as theories on complex vector spaces or algebras (the Schr\"odinger
equation on a Hilbert space and the Heisenberg equation on a
$\mathbb{C}^*$-algebra). It is however convenient to analyse the problem from a
more general perspective, which is manifestly necessary when the character of rays rather
than vector of pures states is taken into account. 
 We hope that a geometrization of quantum mechanics  may be used for a more 
sound theory of Quantum Gravity.
It follows that to `geometrize' quantum theories we should first
describe some fundamental algebraic structures in tensorial terms and then apply this procedure to
describe quantum theories by means of tensorial entities.
The general ideology of this presentation is being elaborated in a book 
provisionally entitled:
 \textit{Geometrical Theory of Classical and Quantum Dynamical Systems}.

We shall proceed  by explaining first what we mean by a geometrical
description of physical systems by considering Newton's
equations and Maxwell's equations, then we consider the usual formulation of 
quantum mechanics and  we finally introduce a tensorial description of the 
algebraic structures emerging in the usual formulation.

\subsection{Newton's equations}

We shall start by considering {\it Newton's equations}, this is a second-order differential equation on some connected
and simply connected configuration space $Q$:
$$\frac{d^2x}{dt^2}= F\left(x,\frac {dx}{dt}\right) .
$$
With this equation we associate a vector field  $\Gamma$ on the tangent bundle $TQ$, or velocity phase space, of  $Q$, say \cite{MSSV},
$$
\Gamma= v\,\pd{}{x}+F(x,v)\pd{}{v}\,.
$$
Having a vector field on a manifold we can use all transformations on $TQ$ to transform it and 
find an easier way to integrate it for instance.     Usually the existence of additional geometrical structures
compatible with the given dynamics will uncover properties of it and will help 
with its integration.   Thus, given a dynamics $\Gamma$ one usually looks for compatible structures, among
them and most noticingly, Poisson brackets.  In other words, one tries to find out if the given dynamics is
Hamiltonian with respect to some (in principle unknown) Poisson brackets.    Poisson brackets are encoded into a
Poisson bivector field, i.e. a $\Gamma$-invariant, contravariant skew-symmetric
2-tensor field $\Lambda$ such that
$\mathcal{L}_{\Gamma}\Lambda=0$. This equation has a clear tensorial meaning. This
bivector field is required to satisfy
$[\Lambda,\Lambda]_{S}=0$, where $[\cdot ,\cdot]_{S}$ is the
Schouten bracket  or, equivalently, the associated Poisson
bracket:
\begin{equation}
\left\{  f_{1},f_{2}\right\}  =\Lambda\left(  df_{1},df_{2}\right)
\end{equation}
should satisfy the Jacobi identity, which is a
quadratic relation.
This condition for $\Lambda$ is a nonlinear partial differential
equation. It may admit no solution, one solution or many solutions. When
it has more than one solution the dynamical system we are
describing is called a bi-Hamiltonian system an exhibits some integrability properties \cite{magri}.  As a matter of fact we
have to distinguish the case of degenerate and non-degenerate
tensors. Whenever the Poisson tensor is non-degenerate one
can define (`modulo' an arbitrary and irrelevant additive
constant) a Hamiltonian function $H$ via
\begin{equation}
\Lambda(\cdot,dH)=\Gamma\,,
\end{equation}
and the dynamics can be
written in Hamiltonian form as:
$${\mathcal{L}}_\Gamma
f=\{f,H\}\,.
$$
 In this case we say that $\Gamma$ defines an inner derivation of the Poisson algebra
 defined by the Poisson bivector field on the space $\mathcal{F}(M)$ of smooth functions on the 
manifold (the tangent  bundle in the case of second-order differential equations). 
If, instead,  $\Lambda$ is degenerate, then $\Gamma$ is still a derivation of the Poisson
algebra but it need not to
be inner and it may define what is called an outer derivation, i.e. it will not be the
image under $\Lambda$ of a 1-form.  When it is the image of a 1-form, the
1-form needs to be closed only on vector fields which define
inner derivations. We shall not insist on these aspects and refer
the reader to the literature \cite{MFLMR}.

When $\Lambda$ is non-degenerate, the condition
$$\Lambda(df,dh)=0\,, \forall f\in \mathcal{F}(M)
$$
implies that the function $h$ is a constant function  and we may
associate a (symplectic) structure $\omega_{\Lambda}$ to $\Lambda$ and the
quadratic condition coming from the Schouten bracket becomes
$d\omega_\Lambda=0$.

In the case of a second-order differential equation, by using $\tau_Q:TQ\to Q$ we may further require the localization property
(i.e., the possibility of measuring simultaneously observables depending only on configuration variables):
$$\Lambda(\tau^*_Qdg_1,\tau^*_Qdg_2)=0\,, \qquad \forall g_1,g_2\in {\mathcal{F}}(Q)\,,
$$
and then we find (see \cite{CIMS}) that there exists a function $L\in {\mathcal{F}}(TQ)$ such that
$$\omega_{\Lambda}=-d\theta_L\,,
$$ 
where $\theta_L = S^*(dL)$ and $S$ denotes the soldering (1,1)-tensor field (or vertical endomorphism) \cite{MFLMR}:
$$S=dx^j\otimes \pd{}{v^j} .
$$
In such a case the dynamics $\Gamma$ may be described in terms of a Lagrangian function $L\in
{\mathcal{F}}(TQ)$ and if we introduce the Liouville vector field \cite{MFLMR}
$$\Delta=v^i\pd{}{v^i}\,,$$
which is the infinitesimal generator of dilations along the fibres, and the energy function $E_L=\Delta L-L$,
 then $(TQ,\omega_\Lambda,E_L)$ is a Hamiltonian system such that $i(\Gamma)\omega_\Lambda=dE_L$.

Then starting from Newton's equations on $Q$ we have defined a second-order
vector field on $TQ$ and if there exists a localizable compatible non-degenerate Poisson tensor $\Lambda$ we have
defined a symplectic structure and a Lagrangian function such that the original dynamics is both a Hamiltonian and
Lagrangian system, completing a geometrization of the original equations of motion.
\medskip

\noindent{\bf Remark:} This approach shows very clearly how we reduce ${\rm
Diff\,}(TQ)$ to  ${\rm Diff\,}(TQ,\Lambda)$ and further to tangent
bundle diffeomorphisms according to Klein's Erlangen programme, i.e. 
we may start with the diffeomorphism group and `break it' to appropriate
subgroups by means of additional structures. These subgroups in general are enough to identify the manifold along with the additional structures.

\subsection{Maxwell's equations}

{\it Maxwell's equations\/} for the electric and magnetic fields,
${\bf E}$ and  ${\bf  B}$,
in empty space and without sources can be written as:
\begin{equation}
\frac d{dt}\matriz{c}{{\bf B}\\{\bf E}}=\matriz{cc}{0&-{\rm rot}\\{\rm
    rot}&0}\matriz{c}{{\bf B}\\{\bf E}}\,,
\end{equation}
which are evolution or dynamical equations on the space of electric and magnetic fields, and:
\begin{equation}\label{constraints}
\nabla\cdot\mathbf{B}=0,\qquad \nabla\cdot\mathbf{E}=0,
\end{equation}
which are {\it constraint\/} equations.   Here again we may rewrite this system
of equations in Hamiltonian form, but the constraints Eqs. (\ref{constraints}) will restrict
the possible Cauchy data we may evolve with the evolutionary
equations.  It is possible to argue, and it is often done, that a Lagrangian description
for these equations is possible by means of a degenerate (or gauge invariant)
Lagrangian written on a bigger carrier space described by vector
potentials $A = ({\bf A}, \phi)$, such that ${\rm rot\,} {\bf A} = {\bf B}$, and ${\bf E} = \dot{\bf A} - \nabla \phi$,   
The introduction of the
vector potential is a way to take into account holonomic
constraints given by div $\mathbf{B}=0$, the constraint on $\mathbf{E}$ being on
`velocities' is a non-holonomic constraint.   Thus we can achieve a geometrization of Maxwell's equations in empty space 
without sources as a non-holonomic degenerate Lagrangian system on the space of vector potentials.
The geometrization of Maxwell's equations in empty space is completed in covariant form when considered as a
theory of connections on a $U(1)$-principal bundle over space-time.

For completeness we
comment on the covariant geometrical formulation of Maxwell's
equations when we consider also sources.
A covariant geometrical formulation \cite{MPT,MT} requires the introduction of the Faraday
2-form
$$
D=E\wedge dt -B\,,
$$
and the Amp\`ere's odd 2-form
$$
G=H\wedge  dt +D\,,
$$
with the odd 3-form
$$J=\rho+j\wedge dt\,.
$$

The equations
$$\left\{\begin{array}{rcl} dF&=&0\,,\\
dG&=&J\,,\end{array}\right.
$$
must be supplemented with phenomenological constitutive equations
$${\mathcal{C}}(F,G,J)=0\,.$$
The constitutive equations are `phenomenological' relations
between the Faraday tensor, the Ampere's tensor and the sources.
They need not be local, i.e.
$${\mathcal{C}}(F,G,J)(x,t)\neq{\mathcal{C}}(F(x,t),G(x,t),J(x,t))\,.$$ 
When they are local, additional geometrical structures may be associated with them.

More general Gauge Theories \cite{BMSS} are also geometrical theories, indeed they can be considered, as Maxwell's equations, theories
of connections in principal bundles, for instance with structural group  $SU(3)\times SU(2)\times
U(1)$.  In this respect also General Relativity may be considered a theory of
(pseudo-) Riemannian connections. Both theories have been considered jointly in
the framework of Kaluza--Klein theories \cite{CLM1,CLM2}.
Once again we have achieved a purely tensorial description of our physical
system.

We will turn now our attention to quantum theories.    

\section{Geometrical description of Quantum Mechanics}

The geometry of quantum mechanics, 
contraryly to what has happened with other physical theories, has played a minor r\^ole after its beginning.   In fact, von Neumann's formulation of quantum mechanics in terms of the theory of Hilbert spaces constituted already a formidable geometrization of quantum mechanics, however, 
further geometrical analysis of the theory was not pursued.

 \subsection{Quantum Mechanics}
 
Quantum mechanics \cite{EM} is usually described in the realm of Hilbert space theory in different ways called `pictures'.   We will concentrate here only on the dynamics of quantum systems, i.e. on the equation describing the evolution of quantum states.    Thus we will associate 
a complex separable Hilbert space $\Hil$ (the set of pure states) with our physical system and the observables of the system are the self-adjoint operators (not necessarily bounded) on $\Hil$.  We will not analyze here other physical aspects of quantum systems like the measure process, etc.

\begin{itemize}

\item Schr\"odinger's picture:  The equation
of motion, Schr\"odinger equation,  is written as:
\begin{equation}\label{schrodinger}
\ii\hbar \pd{\ket{\psi}}t =H\ket{\psi}\,,\qquad \ket{\psi}\in \Hil,\,
\end{equation}
where $H$ is a self-adjoint operator, typically unbounded, on the Hilbert space $\Hil$. 

\item Heisenberg's picture:
 The equations of motion are written as:
\begin{equation}
\ii\hbar \frac{dA}{dt} =[H,A]\,,
\end{equation}
where $A$ and $H$ are self-adjoint operators on $\Hil$.

\item Dirac's interaction picture:
The evolution equation is written now as:
\begin{equation}
\left(\ii\hbar \frac d{dt}U \right)U^{-1}=H\,,
\end{equation}
where $U$ is a unitary operator on $\Hil$.
 This is a generalized Dirac picture written on the group of unitary transformations,
in this sense it is a quantum theory written on a group manifold.

\end{itemize}
All these images can be seen as different realizations in associated vector
bundles of a principal connection (see e.g. \cite{ACP}).

\subsection{Geometrical description of the Schr\"odinger picture}
Previous descriptions of quantum mechanics are given, except for
the generalized Dirac picture, on carrier spaces which are complex linear
spaces. To describe the equations in tensorial terms \cite{CJM} we should
replace linear spaces with real manifolds and linear operators with
tensor fields. We first consider the complex linear space as a real linear space.
Then, to this end we may use our experience in going
from special relativity to general relativity which replaces the
affine Minkowski space with a general Lorentzian manifold. Let us
recall what is done  to go from special relativity described on
some affine space modelled on a Minkowski vector space $V$ with
Minkowskian metric (inner product) $\eta_{\mu\nu}\, x^\mu\, x^\nu$
to a description on a pseudo-Riemannian manifold $M$ by replacing
the Minkowskian inner product with the metric tensor
$g=\eta_{\mu\nu}\, dx^\mu\otimes d x^\nu$.

We may perform a similar trick by formally replacing the inner Hermitean product
$\braket\psi\psi$ on the complex Hilbert space $\Hil$ with the Hermitean tensor
$\braket{d\psi}{d\psi}.$

Let us consider an orthonormal Hilbert basis in $\Hil$,  say $\{\ket{e_j}\mid
j=1,2,\ldots\}$, $\braket{e_j}{e_k}=\delta_{jk}$, and define complex coordinate functions:
\begin{equation}\label{complex_coordinates}
\braket{e_j}{\psi}=z^j=x^j+\ii y^j,\quad \ket{d\psi}=(dz^j)\ket{e_j}.
\end{equation}
Our Hermitean tensor will give rise to:
$$
\braket{d\psi}{d\psi}=(d\bar z^k\otimes dz^j)\,\braket{e_k}{e_j},$$
that written in real coordinates looks like:
$$
\braket{d\psi}{d\psi}=(dx^k\otimes dx^j+dy^k\otimes dy^j)\delta_{kj}+\ii
(dx^k\otimes dy^j-dy^j\otimes dx^k)\delta_{kj}\,.
$$
In this way we obtain a symmetric, Riemannian tensor, and a skew-symmetric symplectic
tensor, i.e.
\begin{equation}
g=dx^k\otimes dx^k+dy^k\otimes dy^k\,,\qquad \omega=dx^k\wedge dy^k\,.\label{thetwostr}
\end{equation}

In a more rigorous way, the Hilbert space  $\Hil$ can be seen as a real space
and then both, a Riemann and a symplectic structure, are defined by:
$$
g(v,w)={\rm Re\,}\langle v,w\rangle\,,\qquad \omega(v,w)={\rm Im\,}\langle v,w\rangle\,.
$$
The 2-form $\omega$ can be shown to be an exact 1-form, i.e. there exists a 1-form
$\theta$ such that $\omega =-d\theta$.
Moreover, there is a complex structure $J$ in 
$\mathcal{H}$ when considered as a real 
space:  the   $\mathbb{R}$-linear 
map corresponding to multiply by the complex 
number $\ii$, $Jv=\ii v$, and therefore such that
$$J^2=- I .$$
This complex structure relates the Riemann and the symplectic structures:
$$g(v_1,v_2)=-\omega(Jv_1,v_2),\qquad  \omega
(v_1,v_2)=g(Jv_1,v_2) ,
$$ 
together with:          
$$g(Jv_1,Jv_2)=g(v_1,v_2)\,.$$

By passing to a contravariant form the structures (\ref{thetwostr}) can be substituted by the
corresponding contravariant tensors:
\begin{equation}\label{GLambda}
G=\pd{}{x^k}\otimes \pd{}{x^k}+\pd{}{y^k}\otimes \pd{}{y^k}\,,\qquad
\Lambda=\pd{}{x^k}\wedge \pd{}{y^k}\,.
\end{equation}
We may associate with the first tensor a bi-differential operator:
\begin{equation}\label{symm}
(f_1,f_2)\equiv G(df_1,df_2)=[[\Delta,f_1],f_2]\,,
\end{equation}
where the Laplacian $\Delta$  is the second-order differential operator:
$$\Delta f=\pd{^2f}{{x^k}^2}+\pd{^2f}{{y^k}^2}\,.
$$
With the skew-symmetric tensor we may associate a Poisson bracket defined as:
\begin{equation}\label{poiss}
\{ f , g \} = \Lambda(df,dg)=\pd{f}{x^k}\, \pd{g}{y^k}-\pd{f}{y^k}\, \pd{g}{x^k}\,.
\end{equation}
To the Schr\"odinger equation, Eq. (\ref{schrodinger}),
we associate the linear equation 
\begin{equation}\label{linear}
\frac {dz^k}{dt}=A^k\,_j\, z^j\,.
\end{equation}
in the coordinate system $z^k$ introduced above, Eq. (\ref{complex_coordinates}).
When $H$ is Hermitean, the matrix $\|A^k\,_j\|$ is skew-Hermitean,
i.e. the infinitesimal generator of a unitary transformation if $H$ defines a self-adjoint operator on $\Hil$.

If we associate a vector field $\Gamma$ with our linear equation,
Eq. (\ref{linear}), as follows:
$$\Gamma =A^k\,_j\,z^j\pd{}{z_k} ,
$$
we find that ${\mathcal{L}}_\Gamma \, \braket{\cdot}{\cdot}=0$, then $\Gamma$
preserves the Hermitean product.  The vector field $\Gamma$ is
at the same time Hamiltonian and Killing, i.e., it preserves both
the symmetric and the skew-symmetric part separately.

We observe that for any Hermitean operator $A$ we may define an evaluation
function which is real valued:
$$
f_A(\psi)=\bra\psi A\ket{\psi} ,
$$
and an expectation value function:
$$
e_A(\psi)=\frac{\bra\psi A\ket{\psi}}{\braket\psi\psi} .
$$
We can compute the symmetric bracket defined before, Eq. (\ref{symm}), for pairs of evaluation functions and
we find that:
$$(f_A,f_B)=G(df_A,df_B)=f_{AB+BA} .
$$
Similarly we can compute the Poisson bracket of two evaluation functions, Eq. (\ref{poiss}), and we get:
$$\{f_A,f_B\}=\Lambda(df_A,df_B)=f_{\ii(AB-BA)}\,.
$$
The function $f_A$ defines a Hamiltonian vector field that with the natural
identification of $T\Hil$ with $\Hil\oplus \Hil$, can be seen to be given by $X_A(\ket{\psi})=-\ii A\ket{\psi}$,
and whose integral curves are the solutions of the equation:
$$\frac d{dt}\ket{\psi}=-\ii A\ket{\psi}\,,
$$
therefore the dynamical evolution corresponding to a given Hamiltonian $H$
is given by Schr\"odinger equation Eq. (\ref{schrodinger}).
Moreover, the expectation value function is such that:
$$(e_A,e_A)=\frac{\bra\psi A^2\ket{\psi}}{\braket\psi\psi}-\left(\frac{\bra\psi
    A\ket{\psi}}{\braket\psi\psi}\right)^2\,,
$$
i.e. the physical interpretation of such a function is clear: $(e_A,e_A)$ is the square of the standard deviation.
\medskip

\noindent\textbf {Remark:}
More precisely, the space of pure states of a quantum system is not associated with a Hilbert space $\Hil$ but to the manifolds of rays
of the Hilbert space $\Hil$, this is to the projective space $\PH$.  For instance, if $\Hil={\mathbb{C}}^2$, we have that
${\mathbb{C}}^2$ is a principal bundle with structural group $\mathbb{C}^*$
and base the projective space $\mathbb{CP}^1$ that can be identified with the two-dimensional sphere $S^2$, that is the space of pure states the system. The bivector field $\Lambda$ on $\mathbb{C}^2$  given by Eq. (\ref{poiss}) is not
projectable on $S^2$, but at each $\ket\psi$ we define two new
tensor fields:
$$\widetilde \Lambda_{\ket{\psi}}=\braket\psi\psi\ \Lambda_{\ket{\psi}}\,,\qquad \widetilde
G_{\ket{\psi}}=\braket\psi\psi\ G_{\ket{\psi}}\,.$$

These two tensor fields are now projectable. They define corresponding
bi-differential operators on $S^2$. 
Note that neither the function $f_A$ is projectable onto the quotient, however $e_A$ is projectable.  
Furthermore, even if the symplectic structure
$\omega$ is projectable, the corresponding potential function $\theta$ is not
projectable and then the projected symplectic form is not exact anymore \cite{bcgb}.

Now we are in the position of defining observables from elements of 
${\mathcal{F}}(S^2,\mathbb{C})$  by requiring that $f\in
{\mathcal{F}}(S^2,\mathbb{R})$
and moreover the Hamiltonian vector field associated with $f$ preserves the
projected symmetric tensor on $S^2$.
Thus, observables are intrinsically defined without any reference to the
original Hilbert space.

One can check that $S^2$ is a Lie-Jordan manifold, i.e., both the Lie and the Jordan
product on observables are mutually compatible and
$$f*g=(f,g)+\ii \{f,g\}-fg
$$
defines a ${\mathbb{C}}^*$-algebra structure when the brackets are
extended to complex-valued functions whose real and imaginary
parts are observables.  We will come back to this point in the following section.

\section{Geometrical description of algebraic structures}

In the previous section we have seen an explicit example where algebraic structures (Hermitian products)
are promoted to geometric objects (Riemannian and symplectic structures) and their properties analyzed from that perspective.
Such procedure is only an instance of a general procedure.   Let us consider a few more
examples of this mechanism of interest not only for quantum theories.

\subsection{Bilinear maps and Frobenius manifolds}
To convey the general ideas we consider real vector spaces.  Let us consider bilinear (or multilinear) maps like:
$$B : V \times V\to V, \quad {\rm or}, \quad b: V\to V^*\,.
$$
It is clear that  $B\in V^*\otimes V^*\otimes V$ and $b\in
V^*\otimes V^*$.
The linear space   $V$ itself can be immersed into ${\mathcal{F}}(V^*)$, by means of the canonical map:
\begin{equation}\label{bidual}
 v\mapsto \widehat v\,, \quad \widehat v(\alpha)=\alpha(v)\,,
\end{equation}
and each vector $v\in V$ can be regarded as a linear map in $V^*$;
hence we may define polynomial functions out of them. 

By introducing a basis for $V$, $\{e_j\}$, and the dual basis $\{ \alpha^k \}$ for
$V^*$, we will have:
$$B = b^l_{jk}\alpha^k\otimes\alpha^j\otimes e_l\,
$$
and now we can promote $B$ to define a tensor field $c_B$ on $V$ by replacing the basis vectors $e_j$ and $\alpha^k$ by 
$dx_j$ and $\partial/\partial x_k$ respectively, this is:
\begin{equation}\label{general}
c_B = b^l_{jk} \, \frac{\partial}{\partial x_j}  \otimes \frac{\partial}{\partial x_k}  \otimes dx_l ,
\end{equation}
(clearly $x_k$ denote linear coordinates on $V^*$ with respect to the basis $\alpha^k$).
A first application of this observation lies in considering the structure constants $b^l_{jk}$, or the tensor $B$,
as the components of an affinely
constant connection $\nabla_B$.   We can also imagine that the tensor $B$ defines a
composition law $\circ$ on the algebra of differential operators by means of:
$$\partial_j\circ \partial_k=b_{jk}\,^l\partial_l .
$$
Then the associativity condition for the product 
is equivalent to the vanishing of the curvature.   Finally, if 
we allow $b_{jk}\,^l$ to depend on the point we get the notion of a Frobenius
manifold, as introduced by Dubrovin \cite{D1,D2,D3,D4}.
Even more, substituting the differential operators $\partial_k$ by their corresponding symbols $p_k$ on the cotangent
bundle $T^*V$,  we can define the quadratic functions: 
$$F_{jk}=p_jp_k-\Gamma_{jk}\,^lp_l\,,
$$
and if $\mathcal{J}$ denotes the ideal generated by them, the associativity condition for the product $\circ$
defined by $B$, becomes
$$\{\mathcal{J}, \mathcal{J} \}\subset \mathcal{J}\,.
$$
This result constitutes the key observation of Magri and Konopelchenko \cite{Ko,KM} to relate Frobenius
manifolds and important hierarchies of completely integrable systems.

\subsection{The Jordan--Lie manifold structure of the space of endomorphisms}
Apart from the association of linear functions on $V^*$ to vectors in $V$ discussed above, Eq. (\ref{bidual}),
to any vector $v\in V$ we can associate the constant vector field $X_v \colon V\to TV $ on $V$ defined by $X_v(w) = (w,v)$.
The Liouville vector field $\Delta\colon v\mapsto (v,v)$ generating infinitesimal
dilations, induces a linear structure on the base manifold from the one on the fiber.

Moreover, by using the identification above of vectors on $V$ with tangent vectors to $TV$, 
any linear transformation $A\colon V\to V$ induces a transformation $T_A$ on tangent vectors:
$$ T_A : TV\to TV, \quad (w,v)\mapsto (w,Av).
$$ 
For instance, $T_I=\Delta$.    In this way the algebra ${\rm End\, }(V)$ is mapped into the algebra of linear
endomorphisms on $TV$ preserving the base.
The map $T_A\mapsto X_A=T_A(\Delta) $ is injective and it therefore allows us to
induce a composition law on vector fields as:
$$X_A\cdot X_C=T_{AC}(\Delta)=X_{AC }\,. $$
If we consider now the Jordan product $\circ$ defined in the space of  
endomorphisms of the linear space $V$ by  $A\circ C=\frac 12 (AC+CA)$,
a similar product is induced on the corresponding vector fields. 
$$X_A\circ X_C=T_{A\circ C}(\Delta)=X_{A\circ C }\,. $$
Now, to any bilinear map $B:V\times V\to V$ we can associate in addition to the tensor field $c_B$ given
by Eq. (\ref{general}) a 2-contravariant tensor field $\tau_B$ on $V^*$ given by $i_\Delta c_B$, or 
 more explicitly:
$$\tau_B(df_1,df_2)(\alpha)=\alpha(B(df_1(\alpha), df_2(\alpha)))\,.
$$
Hence, if we consider the symmetric bilinear composition $B(A,C) = A\circ C$, defined by the Jordan bracket above, the dual space $\mathcal{E}^*$ of the space of endomorphisms $\mathcal{E}$ of the linear space $V$ inherits a symmetric contravariant 2-tensor $\tau_B$ and the corresponding (Jordan) symmetric bracket $(\cdot, \cdot )$ on the algebra of functions $\mathcal{F}(\mathcal{E}^*)$.

The space of endomorphisms $\mathcal{E}$ also carries the Lie algebra bracket:
$$ L(A,C) = \frac 12 [A, C] = \frac 12 (AC - CA) ,$$
inducing the corresponding skew-symmetric contravariant 2-tensor $\tau_L$ on $\mathcal{E}^*$ that defines a Poisson bracket $\{ \cdot, \cdot \} $ on $\mathcal{F}(\mathcal{E}^*)$.
 Thus the space $\mathcal{E}^*$ has the structure of a Jordan-Lie manifold.  The two brackets above are compatible in a trivial way because the two tensors add to the canonical tensor induced by the obvious bilinear map $B_0(A,C) = AC$ on $\mathcal{E}$. 

\medskip

Let us consider the simple example of a complex linear space $V$ of dimension 2.   The space $\mathcal{E}$ of endomorphisms of $V$ has complex dimension 4.  A basis for $\mathcal{E}$ can be chosen as the set of the $2\times 2$ (Hermitean) matrices:
\begin{equation}
\sigma_0=\matriz{cc}{ 1&0\\0&1}\,,\quad \sigma_1=\matriz{cc}{ 0&1\\1&0}\,,\quad \sigma_2=\matriz{cc}{ 0&-\ii\\\ii&0}\,,\quad \sigma_3=\matriz{cc}{ 1&0\\0&-1}\,.
\end{equation}
The corresponding (complex) coordinate functions on matrices are given by:
$$z_\mu(A)=\frac 12 {\rm Tr\,}(\sigma_\mu A)\,, 
$$
and for any $2\times 2$ matrix $A$ we have: $ A = z_\mu  \sigma_\mu$.
Clearly now the skew-symmetric tensor $\tau_L$ becomes:
$$
I=\epsilon_{jkl}\,z_j\,\pd{}{z_k}\wedge\pd{}{z_l},
$$
while the symmetric tensor $\tau_B$ has the form:
$$
R=\pd{}{z_0} \stackrel{\otimes}{_s}
\left(z_j\pd{}{z_j}\right)+z_0\left(\pd{}{z_0}\otimes\pd{}{z_0}+\pd{}{z_1}\otimes\pd{}{z_1}+\pd{}{z_2}\otimes\pd{}{z_2} \right)\,.
$$
Both tensors define a $(1,1)$ tensor field $J$ such that
$J^3=-J$, which is a generalisation of the complex structure.

\subsection{The ${\mathbb{C}}^*$-algebra approach to Quantum mechanics}

The connection of the Schr\"odinger picture with the Heisenberg
picture is provided by the momentum map associated with the
symplectic action of the unitary group on the Hilbert space or the
complex projective Hilbert space of the Schr\"odinger picture. The
inverse connection is provided by the GNS construction which
generalizes to Quantum Mechanics the concept of symplectic
realization of a Poisson manifold.

The linear  transformations preserving both $g$ and $\omega$ as defined in
Eq. (\ref{thetwostr}) constitute the unitary group $\mathcal{U}(\Hil)$. If we
denote by $\mathfrak{u}(\Hil)$ its Lie algebra, the set of skew-Hermitean operators
acting on $\Hil$, and identify the set of all Hermitean operators with the dual
$\mathfrak{u}^*(\Hil)$ via the pairing (in the infinite dimensional case we should restrict to Hilbert-Schmidt operators):
$$ 
\langle A,T\rangle=\frac 12\, {\rm Tr\, }(AT)\,,\qquad A\in \mathfrak{u}^*(\Hil)\,,
T\in \mathfrak{u}(\Hil)\,,
$$
we can consider $\widehat T$ as the linear map associated with $T$, $\widehat
T(A)=\langle A,T\rangle$. A bracket can then be defined as before by: 
$$\{\widehat T_1,\widehat T_2\}=[T_1,T_2]\,\,\widehat{}\ \,,
$$
and similarly a Jordan bracket is introduced by means of:
$$(\widehat{T}_1,\widehat{T}_2)=(T_1T_2+T_2T_1)\,\,\widehat{}\ .
$$
These two brackets are compatible in the sense that they define a Lie--Jordan
algebra in $\mathfrak{u}^*(\Hil)$.
If we consider the inner product on $\mathfrak{u}^*(\Hil)$:
$$\langle A,B\rangle_{\mathfrak{u}^*}=\frac 12\, {\rm Tr\, }(AB)\,,
$$we find that this inner product is preserved by the Hamiltonian vector fields
associated with $\widehat T$ for any $T\in\mathfrak{u}(\Hil)$. These vector fields
are related with corresponding vector fields on $\Hil$, namely,
$$\frac d{dt}(e^{-\ii tA}\ket{\psi})_{t=0}=-\ii\,A\ket{\psi}=X_A(\ket{\psi})\,,
$$
with $\ii\, A\in \mathfrak{u}(\Hil)$. The vector field $X_A$ on $\Hil$ is Hamiltonian
with Hamiltonian function $f_A(\ket{\psi})=\frac 12\, \bra{\psi}A\ket{\psi}$. The momentum
map which relates $X_A$ with the Hamiltonian vector field on
$\mathfrak{u}^*(\Hil)$ associated with $(\ii\, A)\ \widehat{}$ \ \,is given by 
$$\mu : \Hil \to  \mathfrak{u}^*(\Hil), \quad \mu(\ket{\psi}) = \ket\psi\bra\psi .
$$
The symmetric tensor associated with the Jordan bracket:
$$R(d\widehat T_1,d\widehat T_2)=(\widehat T_1,\widehat
T_2)=(T_1T_2+T_2T_1)\,\,\widehat{} \ \,,$$
is a contravariant symmetric 2-tensor as we discussed earlier.  Similarly, the skew-symmetric tensor defined by:
$$I (d\widehat T_1,d\widehat T_2))=\{\widehat T_1,\widehat T_2\} ,$$
is the Poissson tensor associated with the Lie algebra $\mathfrak{u}(\Hil)$. 
These two tensor fields are $\mu$-related with the tensors $G$ and $\Lambda$
defined on $\Hil$, respectively, by Eq. (\ref{GLambda}).

By considering the complex contravariant tensor $R+\ii\, I$ we obtain a tensor
field which allows us to consider the algebra of linear functions on
$\mathfrak{u}^*(\Hil)$ of the form  $\widehat
T+\ii\, \widehat S$, $T,S\in \mathfrak{u}(\Hil)$, 
as a $\mathbb{C}^*$-algebra of complex valued functions.  In this setting the
momentum map relates the Schr\"odinger picture with the Heisenberg picture. To
go from the Heisenberg picture to the Schr\"odinger picture  we consider an
Hermitean realization of the Lie--Jordan algebra on $\mathfrak{u}^*(\Hil)$.
This is a generalisation of the symplectic realisation of the Poisson structure
on $\mathfrak{u}^*(\Hil)$. The existence of these Hermitean realizations for the
Lie--Jordan algebra structure on $\mathfrak{u}^*(\Hil)$, the real part of the
$\mathbb{C}^*$-algebra we are considering, is the essential content of the so
called Gelfand--Naimark--Segal (GNS) construction.  We refer to \cite{CM} for
further details on these interesting aspects.

\end{document}